\documentstyle[aps,prd,epsf,floats]{revtex} 

\def\ss{{\hbox{\boldmath$\sigma$}}}
\def\mm{{\hbox{\boldmath$\mu$}}}

\draft

\begin{document}
\twocolumn[\hsize\textwidth\columnwidth\hsize\csname
@twocolumnfalse\endcsname 

\title{Numerical Toy-Model Calculation of the Nucleon Spin 
Autocorrelation Function\\ in a Supernova Core}

\author{G.~Raffelt} 
\address{Max-Planck-Institut f\"ur Physik 
(Werner-Heisenberg-Institut), 
F\"ohringer Ring 6, 80805 M\"unchen, Germany} 

\author{G.~Sigl} 
\address{Department of Astronomy and Astrophysics, 
The University of Chicago, Chicago, Illinois 60637-1433, USA}

\date{August 27, 1998}

\maketitle

\begin{abstract}
We develop a simple model for the evolution of a nucleon spin in a hot
and dense nuclear medium. A given nucleon is limited to
one-dimensional motion in a distribution of external, spin-dependent
scattering potentials. We calculate the nucleon spin autocorrelation
function numerically for a variety of potential densities and
distributions which are meant to bracket realistic conditions in a
supernova core. For all plausible configurations the width $\Gamma$ of
the spin-density structure function is found to be less than the
temperature $T$. This is in contrast with a naive perturbative
calculation based on the one-pion exchange potential which
overestimates $\Gamma$ and thus suggests a large suppression of the
neutrino opacities by nucleon spin fluctuations.  Our results suggest
that it may be justified to neglect the collisional broadening of the
spin-density structure function for the purpose of estimating the
neutrino opacities in the deep inner core of a supernova.  On the
other hand, we find no indication that processes such as axion or
neutrino pair emission, which depend on nucleon spin fluctuations, are
substantially suppressed beyond the multiple-scattering effect already
discussed in the literature. Aside from these practical conclusions,
our model reveals a number of interesting and unexpected insights.
For example, the spin-relaxation rate saturates with increasing
potential strength only if bound states are not allowed to form by
including a repulsive core. There is no saturation with increasing
density of scattering potentials until localized eigenstates of energy
begin to form.
\end{abstract}

\vskip2.2pc]


\section{Introduction}

The numerical modeling of stellar collapse and supernova explosions
has enjoyed enormous progress which is not quite matched by the poor
development of the microscopic input physics.  This unsatisfactory
state of affairs has recently motivated a number of
authors~\cite{Sawyer89,Burrows,Reddy} to turn their efforts toward a
better understanding of one particular key ingredient, the neutrino
opacities in a hot and dense nuclear medium, for which only crude
prescriptions have thus far been used in supernova codes.  The
emphasis of these works has been to address the composition dependence
of the opacities (hyperons may be important), and to include
many-body effects which range from Pauli phase-space blocking over the
inclusion of modified nucleon dispersion relations to ``screening,''
i.e.\ nucleon-nucleon correlations.
 
We study a further aspect of in-medium neutrino reaction rates which
was not mentioned in these papers, the topic of nucleon
multiple-scattering effects. Their possible importance derives from
the observation that the scattering of neutrinos on nonrelativistic
nucleons is dominated by axial-vector current interactions, i.e.\ it
is primarily a spin-dependent phenomenon.  The nucleon-nucleon
interaction contains a strong spin-dependent piece, leading to a large
nucleon spin-fluctuation rate in a hot nuclear medium. In the
nonrelativistic limit, these spin fluctuations are the sole
microscopic source for the emission of neutrino pairs or axions from a
supernova core or a neutron star~\cite{RS,JKRS}, processes which are
perturbatively described as nucleonic bremsstrahlung.  Moreover, spin
fluctuations inevitably lead to a reduction of the neutrino-nucleon
interaction rate~\cite{Sawyer95,RSS} and to an increase of the energy
transfer rate between the neutrino and nucleon fluids~\cite{JKRS,HR}.

The key quantity for our discussion is the rate $\Gamma$ of
spontaneous nucleon spin fluctuations, or equivalently, the rate by
which a nuclear spin in the medium loses memory of its orientation. Of
course, this process is described by a single constant $\Gamma$ only
if the polarization of a given spin decays exponentially as
$e^{-\Gamma t}$; in general the temporal evolution of the spin
autocorrelation is more complicated. 

A perturbative calculation of, say, neutrino pair emission as a
bremsstrahlung process $NN\to NN\nu\bar\nu$ by virtue of a one-pion
exchange potential implies a value for $\Gamma$ much larger than a
typical temperature $T$ of a supernova core. On the other hand, the
resulting suppression of the neutrino opacities is in blunt conflict
with the large duration of the SN~1987A neutrino signal unless one
discards as background all of the late-time events~\cite{KJR}.  It was
thus concluded that the true $\Gamma$ in a real SN core is probably
not much larger than $T$.  Additional theoretical support for this
empirical conclusion arose from a modified $f$-sum rule for the
spin-density structure function which was derived by one of
us~\cite{Sigl96}.

We presently study a one-dimensional toy model for the nucleon spin
relaxation in a ``background medium'' of spin-dependent potentials. We
believe that our model captures the essential physics of the nucleon
spin evolution, yet is simple enough to allow for an intuitive
understanding and a numerical non-perturbative treatment.  The
conclusion will be that for plausible configurations of density and
potential strengths we always have $\Gamma\alt T$, suggesting that
multiple-scattering effects are indeed not a dominating feature of the
neutrino opacities in the deep inner core of a supernova. Of course,
it also means that the rates for axion or neutrino pair emission are
much smaller than given by a naive bremsstrahlung treatment.
Ironically, then, precisely because perturbation theory is bad in a SN
core implies that a naive calculation of $\nu N$ scattering, ignoring
nucleon spin fluctuations, is a much better approximation than it
deserves!

In Sec.~II we describe and justify the physical structure of
our model. In Sec.~III we discuss our technique for the numerical
calculation of the spin-density structure function. 
Sec.~IV is devoted to numerical results, and in Sec.~V we 
conclude with a discussion and summary. We use natural units,
$\hbar=c=k_B=1$, throughout.


\section{Model}

\subsection{Spin-Density Structure Function}

As a starting point for our model construction we consider a
nonrelativistic single-species medium consisting of nucleons $N$.  In
this case the main quantity of interest for neutrino opacities and
axion emissivities is the positive definite spin-density structure
function
\begin{equation}
S(\omega,{\bf k})=
\int_{-\infty}^{+\infty} dt\,e^{i\omega t}\,R(t,{\bf k}).
\end{equation}
It is the Fourier transform of the normalized spin auto-correlation
function 
\begin{equation}
R(t,{\bf k})\equiv
\frac{4}{3 n_N}
\langle\hat\ss(t,{\bf k})\cdot\hat\ss(0,-{\bf k})\rangle,
\end{equation}
where $\hat\ss(x)=\frac{1}{2}\psi^\dagger(x)\ss\psi(x)$ is the
spin-density operator with $\psi(x)$ the nonrelativistic nucleon field
operator, a Pauli two-spinor, while $\ss$ is the vector of Pauli
matrices. Further, $n_N$ is the nucleon number density and
$\langle\cdots\rangle$ denotes a thermal ensemble average.  The
normalization factor is given by $\langle\hat\ss(0,{\bf
k})\cdot\hat\ss(0,-{\bf k})\rangle =s(s+1)\,n_N$ for all ${\bf k}$
with $s=\frac{1}{2}$ the length of the spin. Therefore, if the
nucleon spins are uncorrelated with each other we have
$R(0,{\bf k})=1$, corresponding to
\begin{equation}
\int_{-\infty}^{+\infty} \frac{d\omega}{2\pi}\,
S(\omega,{\bf k})=1.
\end{equation}
We are primarily interested in the relaxation behavior of $R(t,{\bf
k})$. Therefore, we may equally imagine that possible correlation
effects are absorbed in the definition of $R$ so that $R(0,{\bf k})=1$
can be imposed as a normalization condition.

While $S(\omega,{\bf k})$ is what appears as a scattering kernel in 
the Boltzmann collision equation for neutrinos, it is not the most
practical quantity for a theoretical discussion. It obeys
the principle of detailed balancing,
\begin{equation}
S(-\omega,{\bf k})=e^{-\omega/T}\,S(\omega,{\bf k}),
\end{equation}
and thus is not symmetric. Correspondingly, $R(t,{\bf k})$ has
a nontrivial imaginary part. However, it is enough to consider
a symmetric correlator
\begin{equation}
\bar R(t,{\bf k})\equiv
\frac{3}{4n_N}\,
\frac{\left\langle\hat\ss(t,{\bf k})\cdot\hat\ss(0,-{\bf k})
+\hat\ss(0,{\bf k})\cdot\hat\ss(t,-{\bf k})\right\rangle}{2},
\end{equation}
which is real because $\bar R(t,{\bf k})=\frac{1}{2}[R(t,{\bf
k})+R^*(t,{\bf k})] =\frac{1}{2}[R(t,{\bf k})+R(-t,{\bf k})]$ and thus
is the closest quantum analogue to a classical correlation function.
It is equivalent to a structure function which is even in $\omega$,
\begin{equation}\label{eq:sbar}
\bar S(\omega,{\bf k})= \frac{S(\omega,{\bf k})
+S(-\omega,{\bf k})}{2}.
\end{equation}
The normalization remains
unchanged.  We can always recover the scattering kernel by virtue of
\begin{equation}\label{eq:sbars}
S(\omega,{\bf k})=\frac{2\bar S(\omega,{\bf k})}{1+e^{-\omega/T}},
\end{equation}
bringing out detailed balance explicitly.

\subsection{Long-Wavelength Limit}

In the absence of spin-dependent forces the evolution of the nucleon
spins is trivial. In this case the $\omega$-dependence of $\bar
S(\omega,{\bf k})$ is exclusively caused by nucleon recoils, implying
$\bar S(\omega,0)=2\pi \delta(\omega)$. For a nonvanishing momentum
transfer ${\bf k}$, the width of $\bar S$ is given by the nucleon
recoil and thus is suppressed by the inverse nucleon mass.

The presence of spin-spin interactions changes this picture in that
there will be correlation effects. If we ignore the tensor force as in
Ref.~\cite{Burrows}, the total nucleon spin is still conserved in
binary collisions, allowing for spin dissipation only by
diffusion. While the structure of $\bar S(\omega,{\bf k})$ is now more
complicated, it still has nonvanishing power only for $\omega^2-{\bf
k}^2<0$.  Put another way, the ``screening corrections'' of
Ref.~\cite{Burrows} primarily lead to a reshuffling of the power of
$\bar S(\omega,{\bf k})$, but it continues to have power only where it
has already from recoil effects.

We are primarily interested in the opposite situation, i.e.\ the
effect of the tensor force which allows for local spin dissipation by
coupling the nucleon spins to the orbital motion. It creates power in
$\bar S(\omega,{\bf k})$ for $\omega^2-{\bf k}^2>0$ and, in
particular, has a nonvanishing width even in the long-wavelength limit
${\bf k}\to 0$. Among other consequences this implies that neutrino
pairs and axions can be emitted or absorbed, processes which are
perturbatively described as bremsstrahlung $NN\leftrightarrow
NN\nu\bar\nu$. Without the tensor force such processes do not occur as
can be shown, for example, by an explicit calculation. The
bremsstrahlung picture suggests that $\bar S(\omega,{\bf k})$ depends
only weakly on ${\bf k}$, essentially because momentum conservation
can be achieved among the nucleons alone. Therefore, it is probably
enough to develop an understanding of the behavior of the structure
function in the long-wavelength limit, $\bar S(\omega)=\lim_{{\bf
k}\to0}\bar S(\omega,{\bf k})$.

Our simple final approach is, therefore, to study the single-nucleon
spin autocorrelation function as a proxy for the collisional
broadening of the full dynamical spin-density structure function.

If the effect of collisions on the spin evolution of a given nucleon
can be pictured as a sequence of uncorrelated ``kicks,'' then the
autocorrelation function must decay exponentially,
\begin{equation}\label{eq:rexp}
\bar R(t)=e^{-\Gamma|t|},
\end{equation}
where the quantity $\Gamma$ is what we call the spin-fluctuation or
spin-relaxation rate. If the spin is completely randomized in a given
collision, one can show that $\Gamma$ is identical with the collision
rate $\Gamma_{\rm coll}$.  The structure function corresponding to
Eq.~(\ref{eq:rexp}) is a Lorentzian,
\begin{equation}\label{eq:lor}
\bar S(\omega)=\frac{2\Gamma}{\omega^2+\Gamma^2}.
\end{equation}
A (classical) bremsstrahlung calculation involving single ``kicks''
yields $\bar S(\omega)=2\Gamma/\omega^2$. The Lorentzian structure,
which we obtain directly from a Fourier transform of
Eq.~(\ref{eq:rexp}), is equivalent to a resummation of a sequence of
uncorrelated kicks, i.e.\ it arises from the multiple-scattering
nature of the spin motion caused by a random sequence of collisions.

\subsection{Estimate of {\boldmath $\Gamma$}}

The perturbative estimate of $\Gamma$ is based on a calculation of the
bremsstrahlung process $NN\to NN\nu\bar\nu$ (Appendix~A).
Using a one-pion exchange potential and nondegenerate nucleons it is
found to be approximately $\Gamma/T=1.24\,\rho_{14}\,T_{30}^{-1/2}$
where $\rho_{14}=\rho/10^{14}~{\rm g}~{\rm cm}^{-3}$ and
$T_{30}=T/30~{\rm MeV}$. $\Gamma/T$ of a few is enough to
imply a significant neutrino cross-section suppression (Appendix~B).

The question if the perturbatively estimated $\Gamma$ is realistic can
be addressed by a simple classical estimate. The nucleon spin
relaxation is caused by collisions with other nucleons. The most
effective relaxation is achieved if the spin is completely randomized
in each collision so that $\Gamma=\Gamma_{\rm coll}$. The maximum
collision rate, on the other hand, is estimated by the time it takes a
given nucleon to cross the average distance to the next scattering
center. For nondegenerate nucleons in thermal equilibrium we have
$\frac{1}{2} m_N v^2=\frac{3}{2} T$, leading to a typical velocity of
$v=(3T/m_N)^{1/2}=0.31\,T_{30}^{1/2}$. A typical inverse distance
between nucleons is $n_N^{1/3}=77.3\,\rho_{14}^{1/3}\,{\rm MeV}$.  At
this density the pion-fields produced by the nucleons overlap so that
the average distance between nucleons should, indeed, represent the
distance between scattering events. Therefore, we estimate that the
spin-relaxation rate cannot be larger than
\begin{equation}
\frac{\Gamma_{\rm coll}}{T}\approx
\frac{n_N^{1/3} v}{T}= 0.8\,\rho_{14}^{1/3}
T_{30}^{-1/2},\label{coll3d}
\end{equation}
an estimate which we think is a generous upper limit because a full
spin randomization in each collision is by no means assured.
Therefore, the perturbative result looks too large even for a density
$10^{14}~\rm g~cm^{-3}$, not to mention supranuclear densities,
and the spin relaxation rate is expected to be significantly smaller
than the temperature.

\subsection{One-Dimensional Model} 

The way we performed the classical estimate of the relaxation rate
essentially defines our model. We picture the spin to evolve due to
spin-dependent potentials which the nucleon encounters as it
moves. The spin evolution is thus determined by the potential seen
along the nucleon path which we picture as being one-dimensional---the
nucleon motion in coordinate space should not have a decisive impact
on the (three-dimensional) spin motion. 

We are thus led to ask the following question. If we consider a
nucleon, constrained to one-dimensional motion, how does the
relaxation behavior of the spin depend on the density and strength of
the scattering potentials?  Can the relaxation rate become as large or
even larger than the temperature for configurations which mimic the
situation in a supernova core?

To address these questions we calculate the spin correlation function
for a nucleon which is confined to a one-dimensional box of length
$L$. We place a number of scattering potentials into this box at
random locations, calculate numerically the exact $\bar R(t)$ or $\bar
S(\omega)$, and then take an ensemble average over many
configurations.

A scattering potential at location $x_0$ is taken to interact with the
test nucleon according to a Hamiltonian density of the form
\begin{equation}\label{eq:intham}
H_{\rm int}=\mm(x-x_0)\cdot\hat\ss(x).
\end{equation}
The classical ``magnetic dipole'' is
\begin{equation}
\mm(x)={\bf m}\,V(x)
\end{equation}
with ${\bf m}$ a unit vector pointing in some random direction while
$V(x)$ is a function with the dimension energy. We model it by 
a Gaussian with
\begin{equation}\label{eq:gaussian}
V(x)=V_0\,e^{-(x/b)^2}
\end{equation}
where $b$ is a length scale. 

In order to avoid bound states we will sometimes add a repulsive
diagonal potential so that the total interaction becomes
\begin{equation}
H_{\rm int}=V(x-x_0)\,\,\frac{1}{2}
\psi^\dagger(x)({\bf m}\cdot\ss+1)\psi(x).
\end{equation}
For nucleons polarized in the ${\bf m}$-direction this implies that a
spin-down neutron feels no potential at all while a spin-up neutron
feels twice the (repulsive) potential.

We believe that our approach represents the minimal model that allows
one to study the dependence of the relaxation rate on the potential
strength and density without recourse to perturbative methods.


\section{Numerical Technique}

\subsection{Solving the Schr\"odinger Equation}

For a numerical determination of the spin autocorrelation function we
solve explicitly the one-dimensional Schr\"odinger equation 
\begin{equation}\label{eq:schroedinger}
  \partial_t\left|\psi(x,s_z,t)\right\rangle=
  -i\,H\,\left|\psi(x,s_z,t)\right\rangle,
\end{equation}
where $\psi(x,s_z,t)$ is a nucleon single-particle wavefunction which
depends on the one-dimensional coordinate $x$, the $z$-component
of the spin $s_z=\pm 1/2$, and time $t$. The Hamiltonian is
\begin{equation}
H=-\frac{1}{2m_N}\,\frac{\partial^2}{\partial x^2}+ H_{\rm int}.
\end{equation}
This partial differential equation is solved by discretizing it on a
spatial grid of spacing $a$ so that $x_j=j\,a$, $j=1,\cdots,n_x$,
i.e.\ the box size is $L=(n_x+1)\,a$.  We use Dirichlet boundary
conditions $\psi(0)=\psi(L)=0$, implying that there are $n_x$ grid
points where $\psi(x_j)\not=0$ and thus $n_x$ orthonormal states for
each $s_z=\pm1$.  For the kinetic energy we use the discretized
prescription
\begin{equation}
(H_{\rm kin}\psi)(x_j)=
-\frac{\psi(x_{j+1})+\psi(x_{j-1})-2\psi(x_j)}{2m_Na^2}.
\end{equation}
We then integrate over a time interval $[0,t_{\rm max}]$ in steps of
length $dt$.

In practice, at each time step the complex linear equation
\begin{equation}\label{eq:implicit}
  \left|\psi(t+dt)\right\rangle=\left|\psi(t)\right\rangle+
  \left(\frac{1-i H\,dt/2}{1+iH\,dt/2}-1\right)
  \left|\psi(t+dt)\right\rangle
\end{equation}
is implicitly solved for $\left|\psi(t+dt)\right\rangle$. This
finite-difference representation of $\exp[-i H\,dt]$ is accurate to
second order in $dt$ and unitary.  Equation~(\ref{eq:implicit}) is
employed alternatingly for the $n_x$-dimensional spatial part using
$H_{\rm kin}$ for both spin polarizations, and using the
two-dimensional spin-dependent part $H_{\rm int}$ at all spatial grid
points.

In terms of a complete set of orthonormal functions $|j,s_z\rangle$
the correlation function is given by the expression
\begin{eqnarray}\label{eq:code}
  &&R(t,k)=\nonumber\\
\noalign{\smallskip}
&&\Biggl\langle
  \sum_{{j=1,\ldots, n_x}\atop{s_z=\pm 1/2}}
  \frac{
  \langle j,s_z|e^{-H/T+iHt+ikx}\sigma_z
  e^{-iHt-ikx}
  \sigma_z|j,s_z\rangle}{3Z}
\Biggr\rangle,\nonumber\\
\end{eqnarray}
where the partition function is
\begin{equation}
Z=\sum_{{j=1,\ldots, n_x}\atop{s_z=\pm 1/2}}
  \langle j,s_z|e^{-H/T}|j,s_z\rangle.
\end{equation}
The outer average in Eq.~(\ref{eq:code}) is over many configurations
of the scattering potentials.

The scalar products in Eq.~(\ref{eq:code}) can be written as the
matrix elements ${}_{A,t}\langle j,s_z|
e^{ikx}\sigma_z|j,s_z\rangle_{B,t}$ between the states
$|j,s_z\rangle_{A,t}\equiv e^{-iHt}e^{-H/T}|j,s_z\rangle$ and
$|j,s_z\rangle_{B,t} \equiv e^{-iHt} e^{-ikx}\sigma_z |j,s_z\rangle$.
Put another way, for a complete set of basis functions $|j,s_z\rangle$
we need to solve the Schr\"odinger equation for the states
$e^{-H/T}|j,s_z\rangle$ and $e^{-ikx}\sigma_z |j,s_z\rangle$,
respectively, and then take the matrix element of the operator
$e^{ikx}\sigma_z$. 

The runs presented here will always be for the long-wavelength limit
with $k=0$.  We have verified that $S(\omega,k)$ is indeed nearly
independent of $k$ as long as $k\alt T$. Furthermore, we will always
calculate the symmetric (real) correlator $\bar R(t)= {\rm Re}\,
R(t)$.

\subsection{Long-Time Behavior}
\label{sec:LongTime}

The numerical solutions for $\bar R(t)$ obtained in this way have the
somewhat unintuitive property that they relax to a nonvanishing value
$\lim_{t\to\infty}\bar R(t)=\bar R_\infty>0$. This long-time
behavior is an artifact of the discrete nature of our solution.

This is understood most easily when we write the structure
function in terms of the energy eigenstates $|n\rangle$, 
energy~$\omega_n$, as
\begin{eqnarray}\label{eq:discrete}
S(\omega,{\bf k})&=&
\frac{8\pi}{n_N}\sum_{n,m}
\frac{e^{-\omega_n/T}\left|\left\langle n|\hat\ss(0,{\bf k})
|m\right\rangle\right|^2}{3Z}\nonumber\\
&&\hskip7em{}\times
\delta(\omega+\omega_n-\omega_m),
\end{eqnarray}
where $Z=\sum_n \langle n|n\rangle\, e^{-\omega_n/T}$ is  
the partition function. 

For a finite number of states the weight of the diagonal part of the
double sum ($\omega_n=\omega_m$) does not vanish relative to the
nondiagonal part ($\omega_n\not=\omega_m$). Therefore, $S(\omega,{\bf
k})$ inevitably has a nonvanishing amount of power proportional to
$\delta(\omega)$ even though its spectrum otherwise is continuous
because the ensemble average over different potential configurations
washes out the discrete nature of the energy differences
$\omega_n-\omega_m$ for $n\not=m$. The power of $\delta(\omega)$ shows
up as a positive constant term in the Fourier transform $\bar R(t)$.

\subsection{Short-Time Behavior}
\label{sec:shorttime}

The relaxation behavior of $\bar R(t)$ cannot be exponential at
arbitrarily early times, corresponding to arbitrarily large
frequencies in Fourier space. Very high frequencies correspond to the
(spatial) Fourier components of the scattering potential which is not
arbitrarily hard.  One expects that the (quantum) structure function
has finite moments of the form
\begin{equation}
\int_{-\infty}^{+\infty}\frac{d\omega}{2\pi}\,
\omega^n\,S(\omega)
\end{equation}
for all $n$. (The case $n=1$ is the usual $f$-sum.)  They will exist
if the potential is infinitely often differentiable
everywhere. Assuming that this is the case we conclude that
$S(\omega)$ must fall off faster than any power of $\omega$ and that
the correlation function can be represented as
\begin{equation}
\bar R(t)=1-(\Gamma_2 t)^2+{\cal O}[(\Gamma_2t)^4],
\end{equation}
i.e.\ it has a vanishing derivative at $t=0$.

\subsection{Relaxation Rate}
\label{sec:relaxationrate}

The interpretation of the numerical $\bar R(t)$ would be
straightforward if it were exactly an exponential decay law, which in
general it is not. Therefore, we need a useful prescription to
quantify the ``relaxation rate,'' or equivalently, the ``width'' of
$\bar S(\omega)$. Put another way, we need to define what we mean
with $\Gamma$ for a non-exponential behavior of $\bar R(t)$.

Among many possibilities which differ by a factor of order unity we
find it most natural, for the present purposes, to use the
neutrino-nucleon cross-section reduction as a starting point.  By
virtue of Eq.~(\ref{eq:suppression3}) we can calculate the
cross-section suppression from the numerical $\bar R(t)$, and then
translate it into an equivalent width $\Gamma_{\rm eff}$ of a
Lorentzian that would produce the same cross-section reduction. From
Fig.~\ref{fig:ftau} we conclude that all of the relevant information
of $\bar R(t)$ is contained in the time interval $0\leq t\alt 2/T$.

\subsection{Born Approximation}

For sufficiently weak potentials $V(x)$ a perturbative calculation of
the structure function can be performed in Born approximation which is
a useful test of our numerical code.  For the 3-dimensional case and a
single scattering center such a comparison has been performed in
detail in Ref.~\cite{Sigl97}.  

In one dimension for one scattering
potential of the form Eq.~(\ref{eq:intham}) 
the spin-density structure function in Born approximation in
the long-wavelength limit for $\omega>0$ is
\begin{eqnarray}\label{eq:Born}
  S_{\rm Born}(\omega)&=&\frac{2}{3}\frac{\rho_{\rm s}m_N}{\omega^2}
  \int_0^{\infty}\!\!dp\,f_p\,\frac{\left|\tilde V(p_+)\right|^2
  +\left|\tilde V(p_-)\right|^2}
  {\left(p^2+2m_N\omega\right)^{1/2}}\nonumber\\
&&\hskip3em\bigg/\int_0^{\infty}\!\!dp\,f_p
\end{eqnarray}
where $\tilde V(k)\equiv\int_{-\infty}^{+\infty}dx\,e^{-ikx}V(x)$ is
the Fourier transform of the potential $V(x)$,
$p_\pm=(p^2+2m_N\omega)^{1/2}\pm p$,
$f_p=e^{-p^2/(2m_NT)}$ is the nucleon
occupation number, and $\rho_s$ is the (linear) density of scattering
centers. This structure function is not resummed, i.e.\ it diverges
for $\omega\to 0$, as expected for a bremsstrahlung result without
including multiple scattering. 

For a Gaussian potential of the form Eq.~(\ref{eq:gaussian}) the
Born approximation should be valid when
\begin{equation}\label{eq:cond}
  |V(x)|\ll\frac{1}{m_Nb^2},\quad\hbox{or}\quad
  |V(x)|\ll\frac{p}{m_Nb},\label{Borncond}
\end{equation}
where $p$ is a typical nucleon momentum.

If we take the Gaussian to be very narrow, approaching a
$\delta$-function, then the Fourier transform $\tilde V$ is a constant
and we find
\begin{equation}\label{eq:borndelta}
S_{\rm Born}^0(\omega)=\frac{4(2\pi)^{1/2}}{3}\,
\frac{\rho_s m_N^{1/2}}{T^{1/2}}\,\frac{(V_0 b)^2}{\omega^2}
\int_0^\infty \!\! du\,\frac{e^{-u^2}}{\sqrt{u^2+\omega/T}}.
\end{equation}
This expression diverges at $\omega=0$ because it is a first-order
un-resummed perturbative result. We can still calculate the neutrino
cross-section reduction by virtue of
Eq.~(\ref{eq:suppression2}), which in turn can be translated into an
equivalent width $\Gamma_{\rm Born}^0$ of a Lorentzian which produces
the same cross-section reduction. We find
\begin{equation}\label{eq:borngamma}
\frac{\Gamma_{\rm Born}^0}{T}
=C\,\frac{\rho_s m_N^{1/2}}{T^{3/2}}\,(V_0 b)^2
\end{equation}
with
\begin{eqnarray}
C&=&\frac{4}{3}
\int_0^\infty dx\,\frac{1-(1+x+x^2/6)\,e^{-x}}{(2\pi)^{1/2}A\,x^2}
\int_0^\infty du\,\frac{e^{-u^2}}{\sqrt{u^2+x}}\nonumber\\
&=&\frac{3\pi^2-16}{18\,(2\pi)^{1/2}A}=0.68752,
\end{eqnarray}
where the constant $A$ is defined in Eq.~(\ref{eq:Coefficients}). 

\subsection{Collision Rate}

We finally estimate the rate by which a nucleon in our model
encounters a scattering potential.  If the nucleons are at temperature
$T$, their average kinetic energy in one dimension is $\frac{1}{2} m_N
\langle v^2\rangle =\frac{1}{2} T$ so that $\langle v^2\rangle=T/m_N$.
Taking a Maxwellian velocity distribution tells us that $\langle
|v|\rangle=(2/\pi)^{1/2}\,\langle v^2\rangle^{1/2}$ so that $\langle
|v|\rangle=(2T/\pi m_N)^{1/2}$.  If the average (linear) density of
scatterers in our model is $\rho_s$, then our average collision rate
is $\Gamma_{\rm coll}=\langle |v|\rangle\rho_s$, or
\begin{equation}
\frac{\Gamma_{\rm coll}}{T}=
\left(\frac{2}{\pi T m_N}\right)^{1/2}\,\rho_s.\label{collrate}
\end{equation}
Ultimately, the purpose of our numerical runs is to check if this
estimate can be reached, or even surpassed, for a plausible
configuration.


\section{Numerical Results}

\subsection{Parameters}

We now turn to several series of numerical runs where we have
calculated $\bar R(t)$.  A given run is characterized by the maximum
potential strength $V_0$ (in MeV) and its width $b$ (in fm) according
to Eq.~(\ref{eq:gaussian}) and by whether or not we include a
spin-independent repulsive core (RC) to avoid bound states.  Further
parameters are the temperature $T$ (MeV), the grid spacing $a$ (fm),
the density of scatterers $\rho_s$ (${\rm fm}^{-1}$), the box size $L$
(fm), the total number of scatterers in the box $n_s$, the number of
configurations calculated $n_c$ so that the statistical significance
is proportional to $n_s n_c$, and finally the integration time $t_{\rm
max}$ and the time step $dt$ (both in MeV$^{-1}$).  We usually take
the number of time steps $t_{\rm max}/dt$ to be a power of 2 as
required for a numerical cosine transform of $\bar R(t)$ to determine
$\bar S(\omega)$.

As already indicated, the energy unit is MeV, the time unit
MeV$^{-1}$, and the length unit $1~{\rm fm}=(197~{\rm MeV})^{-1}$. 
Note also that the distance between nucleons scales as 
$n_N^{1/3}=77.3\,\rho_{14}^{1/3}\,{\rm MeV}$ and
$n_N^{-1/3}=2.55\,\rho_{14}^{-1/3}\,{\rm fm}$. We will give
all quantities in these units in the remainder of this
paper. Most runs presented in the following were performed for
$b=1$ and $T=30$, for which Eqs.~(\ref{eq:Born}),
(\ref{eq:suppression2}), and (\ref{eq:limiting}) yield
\begin{equation}
\frac{\Gamma_{\rm Born}}{T}
=3.07\times10^{-4}\,\rho_s\,V_0^2.
\end{equation}
Likewise, the estimated collision rate from Eq.~(\ref{collrate}) is
numerically
\begin{equation}
\frac{\Gamma_{\rm coll}}{T}=
5.13\,\frac{\rho_s}{T^{1/2}}.
\end{equation}
These parameters allow one a first estimate of what to expect in a
given run.

The required CPU-time scales roughly linearly with the step size
$t_{\rm max}/dt$ and with the number of configurations $n_c$, but
quadratically with the number of grid points~$L/a$.

\subsection{Born-Approximation Test}

In order to test our numerical code we compare the (un-resummed)
dynamical structure function in Born approximation of
Eq.~(\ref{eq:Born}) with the one derived from a numerical run. The
characteristic parameters for the run are given in
Table~\ref{tab:Born}.
For these parameters, the condition Eq.~(\ref{eq:cond}) would be
$V_0\lesssim35$. Further, $\Gamma_{\rm Born}/T=7.67\times10^{-4}$ and
$\Gamma_{\rm coll}/T=0.094$, again implying that a perturbative
estimate should be reasonable.

\begin{table}[ht]
\caption{\label{tab:Born}
Parameters for Run {\em Born Test}.}
\smallskip
\begin{tabular}[9]{ccccccccc}
$V_0$ &$b$&RC$^a$&$T$ &$\rho_{\rm s}$ 
&$a$ &$L$  &$t_{\rm  max}$&$dt$\\
\noalign{\vskip3pt\hrule\vskip3pt}
5&1&yes&30&0.1&0.7&100& 81.92&0.005\\
\end{tabular}
$^a$Repulsive Core in the potential.
\end{table}

\begin{figure}[b]
\epsfxsize=8cm
\hbox to\hsize{\hss\epsfbox{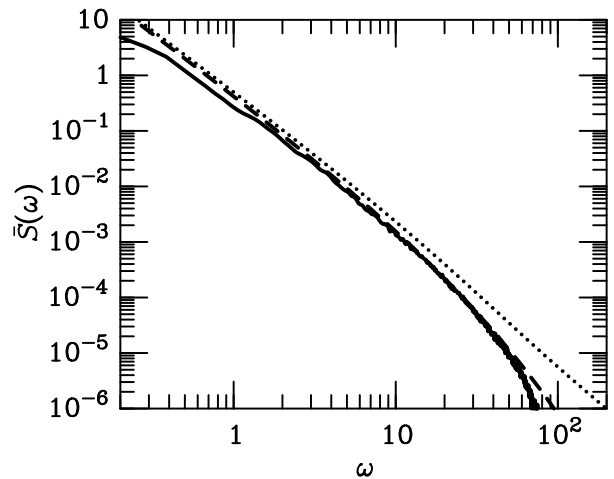}\hss}
\medskip
\caption{\label{fig:Born} Numerical result for $\bar S$ of the {\em
  Born Test} run (solid line), as compared to $\bar S_{\rm Born}$
  (long dashed line).  The dotted line is $\bar S^0_{\rm Born}$,
  corresponding to a $\delta$-function potential.}
\end{figure}

In Fig.~\ref{fig:Born} we compare the numerical $\bar S(\omega)$, with
the Born approximation $\bar S_{\rm Born}(\omega)=S_{\rm
Born}(\omega)\,(1+e^{-\omega/T})/2$ where $S_{\rm Born}(\omega)$ is
from Eq.~(\ref{eq:Born}). In addition, we show the Born approximation
$\bar S^0_{\rm Born}(\omega)$ of Eq.~(\ref{eq:borndelta}) for a
$\delta$-function potential with the same $V_0b$.  The two Born curves
differ only for large $\omega$ where the structure of the potential is
resolved in a collision.  The numerical structure function and the
Born approximation agree very well for intermediate frequencies as
they should.  This agreement is also reflected in the effective spin
fluctuation rate, $\Gamma_{\rm eff}=0.0199$, and $\Gamma_{\rm
Born}=0.0230$. Note that $\Gamma_{\rm Born}^0=0.0487$ is a factor 2.12
larger because a 
$\delta$-function potential produces more power at large
frequencies.

\newpage

\subsection{Variation of the Box Size}

In Sec.~\ref{sec:LongTime} we have remarked that the numerical $\bar
R(t)$ relaxes to a constant value for $t\to\infty$, an effect
attributed to the finite box size of our numerical runs.  In order to
verify that this constant term indeed decreases with increasing box
size, we have performed a series of runs {\em Box Size\/} with the
characteristics summarized in Table~\ref{tab:BoxSize}; there was no
repulsive diagonal potential.

\begin{table}[ht]
\caption{\label{tab:BoxSize}
Common parameters for Series {\em Box Size}.}
\smallskip
\begin{tabular}[6]{cccccccc}
$V_0$ &$b$&RC&$T$   &$a$   &$\rho_{\rm s}$&$t_{\rm  max}$&$dt$\\
\noalign{\vskip3pt\hrule\vskip3pt}
30&1&no&30&1&0.1&32.768&0.002\\
\end{tabular}
%
\bigskip\bigskip
\caption{\label{tab:BoxSizeResults}
Results for Series {\em Box Size}.}
\smallskip
\begin{tabular}[5]{lcccr}
Run$^a$&$n_{\rm s}$&$n_{\rm c}$&$\bar R_\infty$&$1/\bar R_\infty$\\
\noalign{\vskip3pt\hrule\vskip3pt}
B0    &1 & 4096   &0.336&2.98     \\
B1    &2 & 2048   &0.214&4.67     \\
B2    &4 & 1024   &0.157&6.38     \\
B3    &8 & 512    &0.127&7.87     \\
B4    &16& 252    &0.105&9.49     \\ 
B5    &32&  74    &0.104&9.63     \\
B6    &64&  18    &0.090&11.12    \\
\end{tabular}
$^a$Run B$n$ has a box size $L=10\times2^n$.
\end{table}

\begin{figure}[b]
\epsfxsize=8cm
\hbox to\hsize{\hss\epsfbox{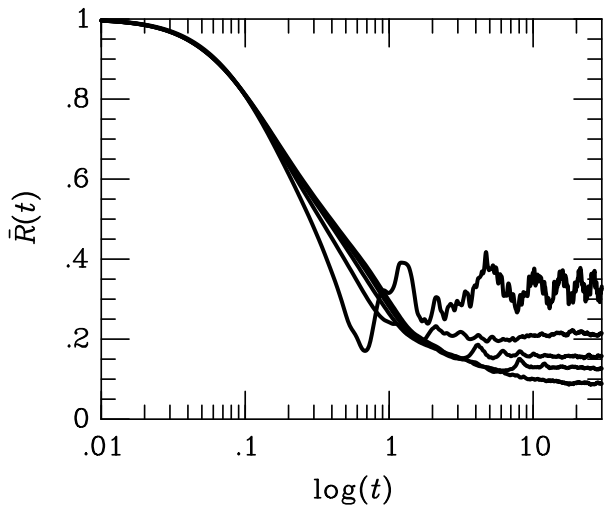}\hss}
\medskip
\caption{\label{fig:box1}Spin autocorrelation function
$\bar R(t)$ for the {\em Box Size} runs B0, B1, B2, B3, B6
(top to bottom).}
\end{figure}

We show the numerical $\bar R(t)$ for several runs in
Fig.~\ref{fig:box1} and summarize the results in
Table~\ref{tab:BoxSizeResults}. We have calculated the limiting
value for $\bar R(t)$ as the average of $\bar R(t)$ for $20\leq t\leq
t_{\rm max}$.  It becomes indeed smaller for an increasing box size,
but only very slowly. Run~B0 with only one scattering center does not
relax to a constant at all; the remaining oscillations of $\bar R(t)$
signal that ``bound-state effects'' in the small box are important.

\begin{figure}[ht]
\epsfxsize=8cm
\hbox to\hsize{\hss\epsfbox{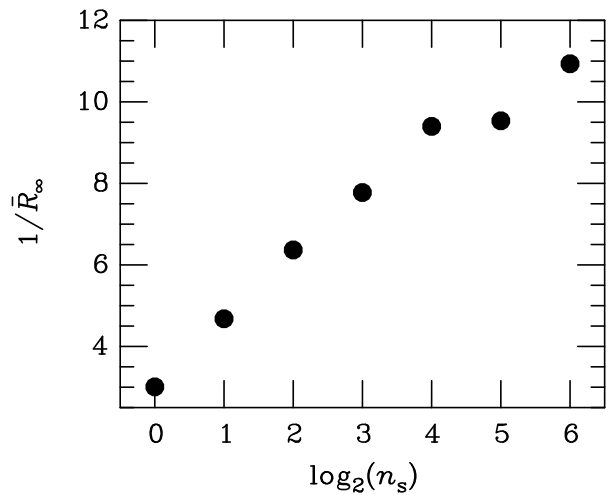}\hss}
\medskip
\caption{\label{fig:rlong} Asymptotic value $\bar R_\infty
=\lim_{t\to\infty}\bar R(t)$ for the {\em Box Size} runs.}
\end{figure}

In Fig.~\ref{fig:rlong} we plot $1/\bar R_\infty$ against
$\log_2(n_s)$, revealing an almost linear behavior. The runs with the
largest box size evidently still have large errors so that their exact
behavior remains numerically uncertain.

For the neutrino cross-section suppression, only the range $t\alt
2/T\approx0.07$ is important (see Appendix~B), where all $\bar R(t)$,
even the one from
run B0, agree perfectly with each other.   Evidently, finite-box
effects are of no relevance if the turn-over to the long-time plateau
occurs at sufficiently late times, i.e.\ at $t\agt2/T$.

\begin{figure}[ht]
\epsfxsize=8cm
\hbox to\hsize{\hss\epsfbox{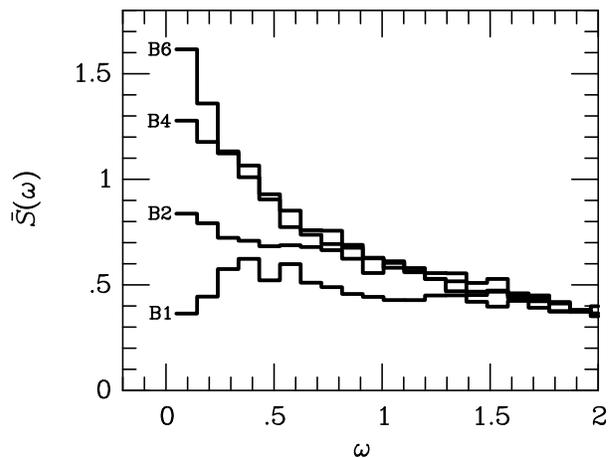}\hss}
\medskip
\caption{\label{fig:box4}Lowest Fourier modes of 
$\bar R(t)$ for some of the {\em Box Size} runs.}
\end{figure}

A similar conclusion is reached by studying the structure function
$\bar S(\omega)$ which differs between different box sizes only for
the few lowest Fourier modes (Fig.~\ref{fig:box4}).  Put another way,
the power which goes into $\bar S(0)$ is missing from the few lowest
$\omega\not=0$ modes, but not from the high modes which
remain unaffected by finite-box effects.

\newpage

\subsection{Variation of the Potential Strength}

When the scattering centers are ``dilute'' so that the nucleons can be
thought of as travelling undisturbed between collisions, the spin
relaxation rate should saturate with increasing potential
strength---the spin can be flipped at most once per encounter. To
demonstrate this effect and to determine the saturation value of the
potential we have performed a series of runs with a Gaussian potential
of varying strength $V_0$. We have included a spin-independent
repulsive potential to avoid bound states which would inevitably
appear as the potential is turned up. The common parameters for this
series {\em Potential Strength} are given in
Table~\ref{tab:PotentialStrength} while parameters and results for
specific runs are summarized in
Table~\ref{tab:PotentialStrengthResults}. The runs are called P$n$
with $n$ the potential strength $V_0$ in~MeV.

\begin{table}[ht]
\caption{\label{tab:PotentialStrength}
Parameters for series {\em Potential Strength}.}
\smallskip
\begin{tabular}[7]{cccccccc}
$b$&RC&$T$   &$a$   &$\rho_{\rm s}$&$n_s$ &$dt$&$\Gamma_{\rm coll}$\\
\noalign{\vskip3pt\hrule\vskip3pt}
1&yes&30&1&0.1&20&0.002&2.81\\
\end{tabular}
\bigskip\bigskip
\caption{\label{tab:PotentialStrengthResults}
Results for series {\em Potential Strength}.}
\smallskip
\begin{tabular}[7]{lcccccc}
Run$^a$&$t_{\rm max}$&$n_{\rm c}$&$\Gamma_{\rm Born}$
&$\Gamma_1$&$\bar R_\infty$&$\Gamma_{\rm eff}$\\
\noalign{\vskip3pt\hrule\vskip3pt}
P2    &65.536& 22&0.00368&--- &--- &0.0033      \\
P4    &32.768& 31&0.0147&--- &--- &0.0135      \\
P8    &16.384&135&0.0589 &0.34&0.10&0.048       \\
P12   &16.384& 89&0.132&0.55&0.15&0.098       \\
P16   &16.384& 40&0.235&0.75&0.17&0.158       \\
P20   & 8.192&107&0.368&0.93&0.20&0.22        \\
P40   & 8.192& 90&1.47&1.23&0.27&0.57        \\
P80   & 8.192& 90&5.89&1.13&0.33&0.95        \\ 
P160  & 8.192& 66&23.5&0.91&0.37&0.96        \\
\end{tabular}
$^a$Run P$n$ has a potential strength $V_0=n$.
\end{table}

\begin{figure}[ht]
\epsfxsize=7.8cm
\hbox to\hsize{\hss\epsfbox{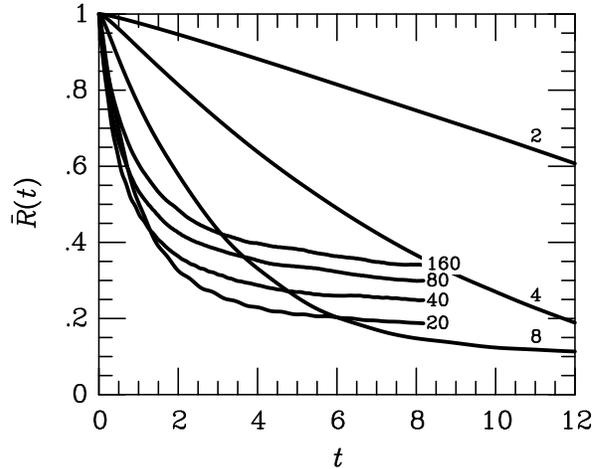}\hss}
\medskip
\caption{\label{fig:pot1} Correlation function $\bar R(t)$ for some
 of the runs of the {\em Potential Strength} series.  The curves are
 marked with $V_0$.}
\end{figure}

\begin{figure}[ht]
\epsfxsize=7.8cm
\hbox to\hsize{\hss\epsfbox{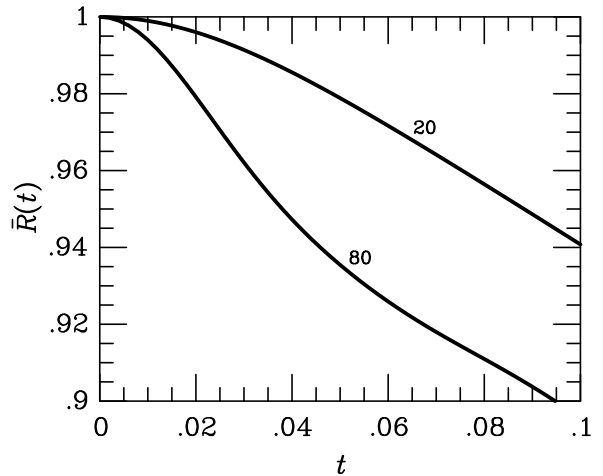}\hss}
\medskip
\caption{\label{fig:pot2}Short-time behavior of
 $\bar R(t)$ for the runs P20 and P80 
 of the {\em Potential Strength} series.}
\end{figure}

In Fig.~\ref{fig:pot1} we show $\bar R(t)$ for most of the runs; the
curve for P16 falls almost exactly on top of P20. The runs P16--P160
show essentially the same relaxation behavior, except that the
long-time value $\bar R_\infty$ increases with increasing potential
strength. This is understood because for large potentials there are
fewer low-lying energy levels that can be occupied by thermal
nucleons, decreasing the effective number of energy eigenstates and
thus increasing the importance of the diagonal term in the
double sum of Eq.~(\ref{eq:discrete}). 

In order to quantify the overall relaxation behavior of the
$\bar R(t)$ curves we approximate them by the ansatz
\begin{equation}
\bar R(t)=\bar R_\infty+(1-\bar R_\infty)\,e^{-\Gamma_1 t}.
\end{equation}
The best-fit parameters are given in
Table~\ref{tab:PotentialStrengthResults}.  For the runs P2 and P4 a
global fit was not meaningful because the overall integration time was
too short.

The overall relaxation rate saturates and the saturation value is
crudely estimated 
by $\Gamma_{\rm coll}$. Still, with increasing potential
the short-time behavior changes significantly for the large
potentials as illustrated in Fig.~\ref{fig:pot2}, i.e.\ an
increasing $V_0$ still implies increasing power at higher modes
of $\bar S(\omega)$. However, $\bar R(t)$ falls off more slowly
than a global exponential because of its vanishing derivative
at $t=0$. Hence the neutrino-cross
section reduction is smaller than expected from 
the relaxation rate $\Gamma_1$, and saturates at larger values
for $V_0$.  The entries in 
Table~\ref{tab:PotentialStrengthResults} for $\Gamma_{\rm eff}$
confirm this trend; the saturation value for $\Gamma_{\rm eff}$
is very similar to that of $\Gamma_1$, i.e.\ about 1~MeV, but
the potential strength at which this value is reached is
$V_0\simeq16$ for $\Gamma_1$, as compared to $V_0\simeq80$
for $\Gamma_{\rm eff}$.

We may also compare $\Gamma_{\rm eff}$ with the simple Born estimate.
Evidently, the Born approximation gives us a reasonable estimate of
$\Gamma_{\rm eff}$ and thus of the cross-section suppression up to
$V_0\alt20$, but substantially overestimates $\Gamma_{\rm eff}$
for higher potential strengths.

\begin{figure}[ht]
\epsfxsize=8cm
\hbox to\hsize{\hss\epsfbox{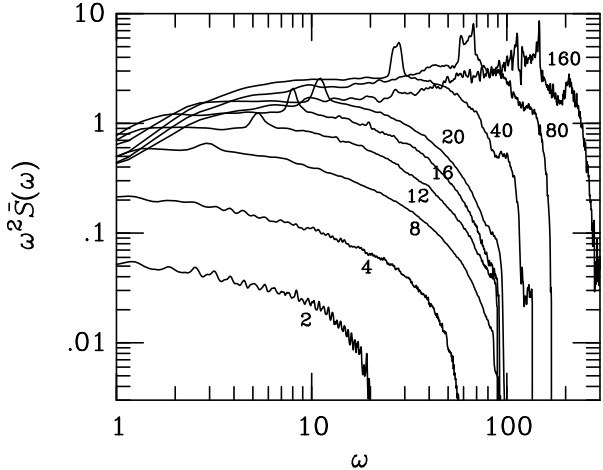}\hss}
\medskip
\caption{\label{fig:pot3}Dynamical structure function for
the runs of the {\em Potential Strength} series.}
\end{figure}

Finally, we show in Fig.~\ref{fig:pot3} the dynamical structure
function $\bar S(\omega)$ obtained from cosine-transforming
$\bar R(t)$. We actually show $\omega^2 \bar S(\omega)$, a quantity
that would asymptotically reach $2\Gamma$ for large $\omega$
if $\bar S(\omega)$ were a Lorentzian. For $V_0\agt12$ the
power spectra develop a conspicuous resonance feature.

\begin{figure}[ht]
\epsfxsize=8cm
\hbox to\hsize{\hss\epsfbox{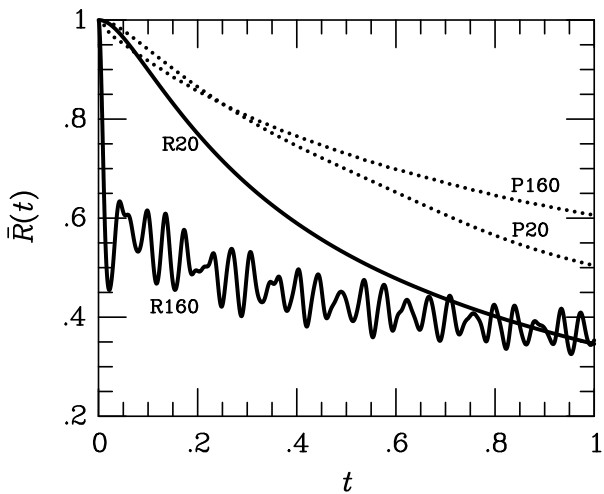}\hss}
\medskip
\caption{\label{fig:pot5}Correlation functions with (dotted) and
without (solid) a repulsive spin-independent core for the potential.}
\end{figure}

Such resonance effects are even more important if we use a potential
without a spin-independent repulsive term. This is illustrated by a
similar series of runs with common characteristics summarized in
Table~\ref{tab:Pchar} and run-specific information in
Table~\ref{tab:Presults}. Of course, in the perturbative regime a
spin-independent potential has no effect on the spin relaxation rate
so that the presence or absence of this term begins to show up only
when the impact of the potentials is no longer small.  The impact of
the repulsive core on $\bar R(t)$ is illustrated by
Fig.~\ref{fig:pot5} for $V_0=20$ and $160$. The difference in
$\Gamma_{\rm eff}$ is still $\simeq10\%$ at $V_0=5$.

\begin{table}[ht]
\caption{\label{tab:Pchar}
Parameters for series {\em Resonances}.}
\smallskip
\begin{tabular}[6]{cccccc}
$b$&RC &$T$   &$a$   &$\rho_s$&$\Gamma_{\rm coll}$\\
\noalign{\vskip3pt\hrule\vskip3pt}
1&no&30&1&0.1&2.81\\
\end{tabular}
\bigskip
\bigskip
\caption{\label{tab:Presults}
Results for series {\em Resonances}.}
\smallskip
\begin{tabular}[7]{lcccccc}
Run$^a$&$t_{\rm max}$&$10^3dt$&$n_{\rm s}$&$n_{\rm c}$&
$\Gamma_{\rm Born}$&$\Gamma_{\rm eff}$\\
\noalign{\vskip3pt\hrule\vskip3pt}
R5     &65.536&2    & 20  &  23 &0.0230&0.023\\
R10    &32.768&2    & 20  &  54 &0.0920&0.091\\
R20    &16.384&2    & 20  &  92 &0.368& 0.37\\
R30    &16.384&2    & 10  & 240 &0.828& 0.83\\
R40    & 8.192&1    &  5  & 932 &1.47& 1.51\\
R60    & 8.192&0.5  &  5  & 265 &3.31& 3.5 \\
R80    & 4.096&0.25 &  5  & 280 &5.89& 6.3 \\
R120   & 4.096&0.25 &  5  & 301 &13.2&14.8 \\ 
R160   & 2.048&0.125&  5  & 213 &23.5&28.5 \\
\end{tabular}
$^a$Run R$n$ has a potential strength $V_0=n$. 
\end{table}

\begin{figure}[ht]
\epsfxsize=8cm
\hbox to\hsize{\hss\epsfbox{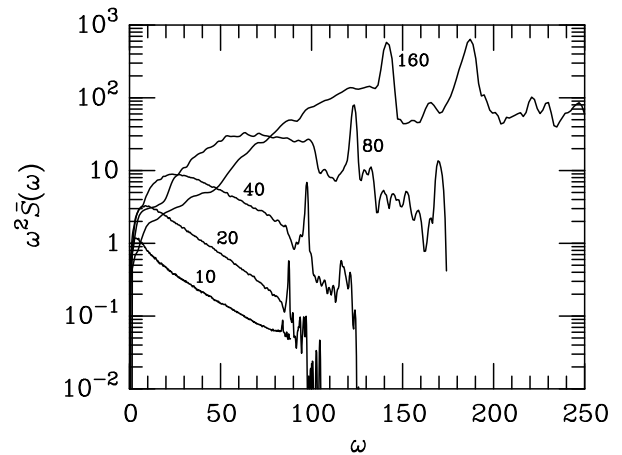}\hss}
\medskip
\caption{\label{fig:pot4}Dynamical structure function for
some of the runs of the {\em Resonance} series.}
\end{figure}

In Fig.~\ref{fig:pot4} we show the dynamical structure function for
some of these runs. When bound states are important, they have far
more power at large frequencies than the previous runs with a
repulsive core.  Likewise, the effective relaxation rate $\Gamma_{\rm
eff}$ increases with increasing $V_0$; there is no saturation.
Indeed, both $\Gamma_{\rm Born}$ and $\Gamma_{\rm eff}$ scale as
$V_0^2$. For all runs $\Gamma_{\rm Born}/\Gamma_{\rm eff} \approx1$,
and thus the Born approximation is surprisingly good,
despite being not expected to be applicable
according to Eq.~(\ref{Borncond}).
For $V_0=160$ we reach $\Gamma_{\rm eff}/T=1$, but of course, such
large potentials are far beyond what appears plausible for a supernova
core where bound-state effects probably do not dominate.

\newpage

\subsection{Variation of the Potential Width}

In order to check to which degree the saturation value of $\Gamma_{\rm
eff}$ for large potential strengths $V_0$ depends on the width $b$
of the potential, we have calculated
$\Gamma_{\rm eff}$ for a series of combination of $V_0$ and $b$.  For
these runs the integration time was taken very short, just long enough
to determine $\Gamma_{\rm eff}$, but not long enough to calculate
a detailed power-spectrum or other global information.  The common
parameters for this series {\em Potential Width} are given in
Table~\ref{tab:PotentialWidth} while the $\Gamma_{\rm eff}$-values are
shown in Table~\ref{tab:PotentialWidthResults}.

\begin{table}[ht]
\caption{\label{tab:PotentialWidth}
Parameters for series {\em Potential Width}.}
\smallskip
\begin{tabular}[9]{ccccccccccc}
RC&$T$   &$a$   &$\rho_{\rm s}$&$n_s$& $L$ & $dt$ & $t_{\rm max}$
&$n_c$&$\Gamma_{\rm coll}$\\
\noalign{\vskip3pt\hrule\vskip3pt}
yes&30&0.5&0.1&5&50&0.001&0.1&500&2.81\\
\end{tabular}
\bigskip\bigskip
\caption{\label{tab:PotentialWidthResults}
$\Gamma_{\rm eff}$ for series {\em Potential Width}.}
\smallskip
\begin{tabular}[9]{rcccccccc}
  &\multicolumn{8}{c}{$V_0$}\\
\noalign{\vskip2pt}
&25&50&75&100&125&150&175&200\\
\noalign{\vskip2pt}
\hline
\noalign{\vskip2pt}
$b=0.5$&0.13&0.33&0.47&0.55&0.57&0.58&0.54&0.52\\
   1.0 &0.31&0.73&0.96&1.08&1.07&1.03&0.95&0.93\\  
   1.5 &0.50&1.12&1.48&1.63&1.59&1.54&1.46&1.39\\ 
   2.0 &0.68&1.53&2.03&2.19&2.12&2.11&1.98&1.90\\ 
\end{tabular}
\end{table}

The saturation value of $\Gamma_{\rm eff}$ increases with $b$, but
for widths which represent the range
of the nuclear force, it never reaches the
``collision limit'' $\Gamma_{\rm coll}$.

\subsection{Variation of the Density of Scatterers}

\begin{figure}[ht]
\epsfxsize=8cm
\hbox to\hsize{\hss\epsfbox{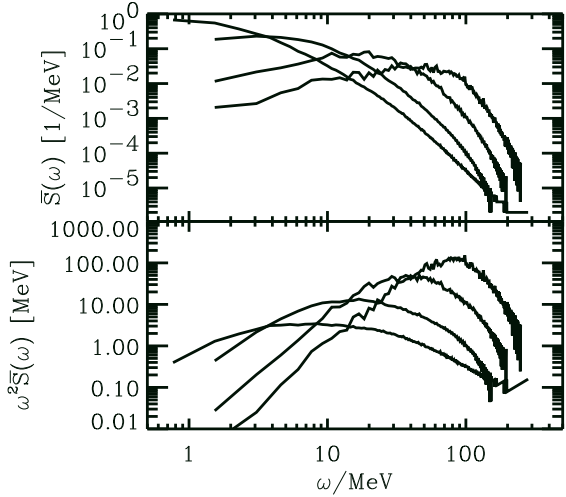}\hss}
\medskip
\caption{\label{fig:wd30a} $\bar S(\omega)$ (upper panel) and
$\omega^2\bar S(\omega)$ (lower panel) for some of the runs of the
{\em Density} series with repulsive core.  In decreasing order at the
left end, the lines represent the runs D10, D40, D200, and D600.}
\end{figure}

\begin{figure}[ht]
\epsfxsize=8cm
\hbox to\hsize{\hss\epsfbox{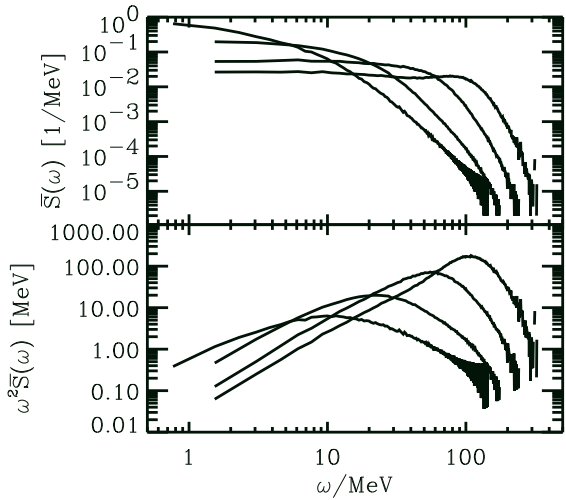}\hss}
\medskip
\caption{\label{fig:wd30}
Same as Fig.~\ref{fig:wd30a}, but for the {\em Density} series
without repulsive core., i.e., the runs E10, E40, E200, and E600.}
\end{figure}

As a next test we have varied the density of scatterers, keeping the
potential strength fixed at $V_0=20$ where the efficiency of
spin flips saturates according to the results of the previous section.
We have performed two series of runs, one (D) where the potentials
include a repulsive core to avoid bound states, and one
(E) without this term. The common parameters for both series are
given in Table~\ref{tab:density}, while parameters and
results specific to the series D and E are given in
Tables~\ref{tab:Dresults} and~\ref{tab:Eresults}, respectively,
and the dynamical spin structure functions for some of these
cases are shown in Figs.~\ref{fig:wd30a} and~\ref{fig:wd30}.
The runs are named D$n$ or E$n$ where $n$ is the number of scatterers
in a given configuration.

\begin{table}[ht]
\caption{\label{tab:density}
Parameters for series {\em Density}.}
\smallskip
\begin{tabular}[5]{ccccccc}
$V_0$ &$b$ &$T$   &$a$   &$L$&$t_{\rm  max}$&$dt$\\
\noalign{\vskip3pt\hrule\vskip3pt}
20&1&30&0.5&50&4.096&0.001\\
\end{tabular}
\bigskip\medskip
\caption{\label{tab:Dresults}
Results for {\em Density\/} (repulsive core).}
\smallskip
\begin{tabular}[6]{lccccc}
Run&$\rho_{\rm s}$&$n_{\rm c}$&$\Gamma_{\rm coll}$
&$\Gamma_{\rm Born}$&$\Gamma_{\rm eff}$\\
\noalign{\vskip3pt\hrule\vskip3pt}
D10 &0.2 & 166 &5.6 &0.736&0.46   \\
D20 &0.4 & 164 &11.2 &1.47&0.90   \\
D40 &0.8 & 164 &22.5 &2.94&1.76   \\
D60 &1.2 & 164 &33.7 &4.72&2.53   \\ 
D100&2.0 & 146 &56.2 &7.36&4.1    \\
D200&4.0 & 145 &112 &14.7&8.1    \\
D600&12.0& 340 &337  &44.2&27.9      \\
\end{tabular}
\bigskip\medskip
\caption{\label{tab:Eresults}
Results for {\em Density\/} (no repulsive core).}
\smallskip
\begin{tabular}[6]{lccccc}
Run&$\rho_{\rm s}$&$n_{\rm c}$&$\Gamma_{\rm coll}$
&$\Gamma_{\rm Born}$&$\Gamma_{\rm eff}$\\
\noalign{\vskip3pt\hrule\vskip3pt}
E1$^a$&0.02& 56&0.56 &0.0736&0.067 \\
E3$^b$&0.06& 60&1.69 &0.221&0.234 \\
E5$^b$&0.10& 54&2.81 &0.368&0.38  \\
E7$^b$&0.14& 60&3.93 &0.515&0.51  \\
E10   &0.2 &166&5.6 &0.736&0.72  \\
E20   &0.4 &166&11.2 &1.47&1.39  \\
E40   &0.8 &166&22.5 &2.94&2.69  \\
E60   &1.2 &166&33.7 &4.42&4.0   \\ 
E100  &2.0 &154&56.2 &7.36&6.4   \\
E200  &4.0 &163&112 &14.7&11.6  \\
E600  &12.0&250&337 &44.2&40.0  \\
\end{tabular}
$^a$$t_{\rm max}=4\times4.096$ and $dt=0.002$.\\
$^b$$t_{\rm max}=2\times4.096$.
\bigskip\bigskip
\caption{\label{tab:DVresults}
Results for {\em Density\/} (repulsive core, $V_0=80$).}
\smallskip
\begin{tabular}[6]{lccccc}
Run&$\rho_{\rm s}$&$n_{\rm c}$&$\Gamma_{\rm coll}$
&$\Gamma_{\rm Born}$&$\Gamma_{\rm eff}$\\
\noalign{\vskip3pt\hrule\vskip3pt}
DV3$^a$  &0.06 & 55 &1.69 &3.53&0.609 \\
DV10$^b$ &0.2 & 40 &5.6 &11.8&1.95 \\
DV40$^c$ &0.8 & 47 &22.5 &47.1&7.89 \\
DV200$^c$&4.0 & 425 &112 &235&54.3 \\
DV600$^c$&12.0& 108 &337  &706&104 \\
\end{tabular}
$^a$$t_{\rm max}=2\times4.096$.\\
$^b$$dt=0.0005$.\\
$^c$$t_{\rm max}=4.096/2$ and $dt=0.0005$.
\end{table}

\begin{figure}[ht]
\epsfxsize=8cm
\hbox to\hsize{\hss\epsfbox{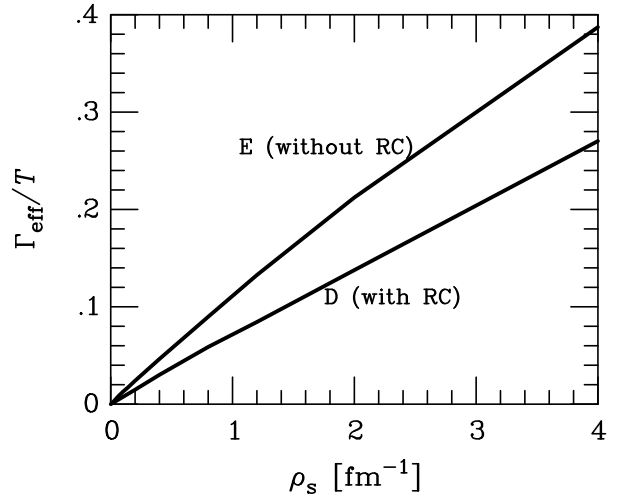}\hss}
\medskip
\caption{\label{fig:dens2}Effective relaxation rate for
{\em Density} runs.}
\end{figure}

The effective relaxation rate grows linearly with density
(Fig.~\ref{fig:dens2}) even for densities so large that the potentials
significantly overlap with each other. We have observed that this
holds for any potential strength as long as most of the energy
eigenfunctions are not localized. The absence of saturation can also
be seen from the dynamical structure functions themselves,
Figs.~\ref{fig:wd30a} and~\ref{fig:wd30}. Note that at high densities
the structure functions exhibit the flat behavior at low frequencies
expected from the classical picture, Eq.~(\ref{eq:lor}), in the case
without a repulsive core, whereas in the presence of a repulsive core
they increase with frequency.

As soon as the potentials are strong enough and/or the density is high
enough for a significant fraction of the energy eigenfunctions to
become localized we observe a saturation in $\rho_s$, as demonstrated
in Table~\ref{tab:DVresults} which lists results for potentials of
strength $V_0=80$ and with a repulsive core. This can be understood
qualitatively by observing that $\left|\left\langle n|\hat\ss(0,{\bf
k})|n\right\rangle\right|^2$ is independent of the box size $L$ for a
localized eigenstate $|n\rangle$, as opposed to decreasing with $L$
for a non-localized state, and thus gives rise to an asymptotic
non-vanishing $\bar R_\infty$ in the continuum limit $L\to\infty$,
preventing the scattering cross section $\langle\sigma\rangle$ to
decrease further [see Eq.~(\ref{eq:suppression3})]. At the same time,
we expect the test-particle approximation adopted in the present work
to break down in the regime with a significant fraction of localized
states.

Outside this regime, the highest effective spin fluctuation rate and
the strongest scattering cross section suppression we found were
$\Gamma_{\rm eff}/T\simeq2$ and $\langle\sigma\rangle/\sigma_0
\simeq0.5$ for $V_0\simeq80$, $\rho_s\simeq4$, about a factor 2
smaller than $\Gamma_{\rm coll}$, see Table~\ref{tab:DVresults}.

Finally, to get a better overview of the saturation behavior
for large potential strengths $V_0$ in the
presence of a repulsive core, we have calculated $\Gamma_{\rm eff}$
for a series of combination of $V_0$ and $\rho_s$.  Again, for these
runs the integration time was taken very short.  The common parameters
for this series {\em Strength-Density} are given in
Table~\ref{tab:StrengthDensity} while the $\Gamma_{\rm eff}$-values
are shown in Table~\ref{tab:StrengthDensityResults}.

\begin{table}[ht]
\caption{\label{tab:StrengthDensity}
Parameters for series {\em Strength-Density}.}
\smallskip
\begin{tabular}[8]{cccccccc}
$b$ &RC&$T$   &$a$   & $L$ & $dt$ & $t_{\rm max}$&$n_c$\\
\noalign{\vskip3pt\hrule\vskip3pt}
1&yes&30&1&50&0.001&0.1&500\\
\end{tabular}
\bigskip\bigskip
\caption{\label{tab:StrengthDensityResults}
$\Gamma_{\rm eff}$ for series {\em Strength-Density}.}
\smallskip
\begin{tabular}[9]{rrrrrrr}
  &&\multicolumn{5}{c}{$V_0$}\\
\noalign{\vskip2pt}
&&25&50&75&100&150\\
\noalign{\vskip2pt}
\hline
\noalign{\vskip2pt}
$\rho_s=0.1$&$\Gamma_{\rm coll}=2.81$&0.32&0.70&0.92&1.03&0.99\\
        0.5 &                   14.0 &1.59& 3.6& 4.7& 5.4& 5.2\\
        1.0 &                   28.1 & 3.1& 7.1& 9.7&10.9&11.4\\ 
        2.0 &                   56.2 & 6.2&14.5&22. &27. & 38.\\
        4.0 &                  112.4 &12.8&33. &50. &64. & 87.\\ 
\end{tabular}
\end{table}

Again, even for extreme combinations of density and potential 
strength, does $\Gamma_{\rm eff}$ stay safely below
the collisional estimate $\Gamma_{\rm coll}$. For
combinations which may plausibly mimic nuclear matter,
$\Gamma_{\rm eff}$ always stays below $T$.


\newpage

\section{Discussion and Summary}

We have numerically studied the relaxation behavior of a single
nucleon which moves in a one-dimensional box, filled with a certain
density of spin-dependent scatterers. This simple model is meant to
mimic the collisional broadening of the dynamical nucleon spin-density
structure function in a supernova core.

In most runs, the potential was modelled as a Gaussian with a width of
$1~{\rm fm}$ as suggested by the range of the one-pion exchange force
which is the primary cause for nucleon spin fluctuations. However, the
exact range of the potential has no significant impact on our
conclusion.

In order to quantify the spin relaxation behavior by a single number
we have chosen an effective spin relaxation rate $\Gamma_{\rm eff}$
which is defined as the width of a Lorentzian structure function which
produces the same neutrino-nucleon cross section suppression.  For
sufficiently weak potentials, our numerical runs reproduce the
perturbative value in Born-approximation.

When the potential has a repulsive core such that bound states cannot
form, the spin-relaxation rate saturates with increasing potential
strength at a value which is a factor 2--3 smaller than one would have
expected from a classical analogy where one pictures the spin as being
randomized each time the nucleon encounters a scattering center. The
saturation value for the potential strength roughly coincides with the
value when the Born approximation begins to break down, and where
bound states begin to form in the absence of the repulsive core. For
the range $1~{\rm fm}$ of our Gaussian potential this is at a
potential strength of around 20~MeV which also is typical for the
spin-interaction energy between two nucleons. 

Furthermore, we observed that the spin-relaxation rate is roughly
linear in the density of scatterers. This is expected in the classical
picture, but our simulations show that it even holds if the potentials
overlap substantially.  Saturation is only observed for densities high
and/or potentials strong enough for the formation of a substantial
fraction of localized energy eigenstates. However, in this regime, a
strong correlation of neighboring nucleon spins is expected, and the
test-particle approximation adopted in our work is no longer
meaningful.

When the scattering potentials lead to bound states (no repulsive
core), there is no saturation of $\Gamma_{\rm eff}$, not with
increasing potential strength and not with an increasing density of
scatterers. Instead, the spin-relaxation rate follows the Born
approximation amazingly closely, despite the fact that it is not
applicable. However, the test-particle approximation again is not
viable as a model for a nuclear medium because the bound states which
dominate in this regime signal the importance of
spin-spin-correlations.

In any case, we expect the case of repulsive potentials without
strongly bound states to be more appropriate as a model for the
conditions in a supernova core because of the relevant properties of
nuclear matter.  In this situation, the value $\Gamma_{\rm coll}$
derived from a classical picture provides an upper limit for the
spin-fluctuation rate which saturates for potential strengths expected
from nuclear interactions within about a factor 2, and captures the
linear dependence on the one-dimensional density in the regime where
spatial spin correlations are small and the test-particle
approximation is applicable. 

For combinations of density and potential strength which do not look
like absurd overestimates of what is appropriate for nuclear matter,
the spin-fluctuation rate is always safely below the temperature, and
the scattering cross-section suppression is at most a few ten percent.
The Born approximation, in contrast, significantly overestimates the
spin-fluctuation rate and predicts far too large cross-section
reductions.

Naturally, the conclusions from our one-dimensional model do
not directly carry over to three dimensions, but can only be taken as
an indication of what one might expect.  If the classical collision
rate for three dimensions embodied by our Eq.~(\ref{coll3d}) is taken
as an upper limit for the spin relaxation rate, then again a naive
Born-approximation calculation is a significant overestimate, and the
true collisional broadening of $S(\omega,k)$ is too small to cause a
significant cross-section reduction.

Our calculations also indicate that the Born approximation, while
being a significant overestimate of the spin-relaxation rate, it is
not an absurd overestimate. For the relevant conditions it may
overestimate $\Gamma$ by a factor of a few, but not by orders of
magnitude.  Therefore, we do not expect that quantities like the axion
or neutrino-pair emission rate, which depend on the collisional
broadening of $S(\omega,k)$, suffer much of an additional suppression
beyond those discussed in Ref.~\cite{JKRS}.

From our simple toy model we find no evidence for extreme deviations
from the overall picture developed in Ref.~\cite{JKRS,Sigl96}, which
may be summarized by saying that spin fluctuations look small enough
to avoid excessive neutrino opacity reductions, but large enough to
avoid huge reductions of the axion or neutrino pair emissivities.

Clearly, more accurate and detailed calculations should be performed
in three dimensions with realistic nucleon interaction potentials.
Furthermore, a treatment of the high-density limit should go beyond
the test-particle approximation and apply a full quantum Monte Carlo
approach~\cite{koonin97} to finite temperature and nuclear matter.


\section*{Acknowledgments}

In Munich, this work was supported, in part, by the Deutsche
Forschungsgemeinschaft under grant No.\ SFB-375, at the University of
Chicago by the DoE, NSF, and NASA.


\newpage

\appendix

\section{Perturbative~Estimate~of~{\boldmath $\Gamma$}}

\begin{figure}[ht]
\epsfxsize=8cm
\hbox to\hsize{\hfill\epsfbox{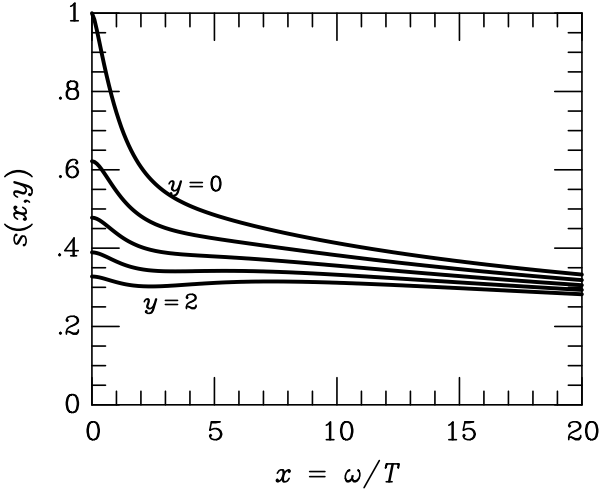}\hfill}
\smallskip
\caption{\label{fig:snd}
Dimensionless structure function according to
Eq.~(\protect\ref{eq:snd}) for $y=0,\ldots,2$ in steps of 0.5.}
\end{figure}

From a perturbative calculation of the bremsstrahlung process $NN\to
NN\nu\bar\nu$ one can easily extract the (unresummed) dynamical
structure function. For nondegenerate nucleons and on the basis of a
one-pion exchange potential it has been evaluated in Ref.~\cite{RS};
even for the conditions of a supernova core this potential should be a
reasonable representation of the tensor force~\cite{HR}. The result
was written in the form
\begin{equation}
\bar S^{(1)}(\omega)=\frac{\Gamma_\sigma}{\omega^2}\,\bar s(\omega/T),
\end{equation}
where $\bar s(x)$ is a slowly varying 
dimensionless function of order 1, normalized such that
$\bar s(0)=1$. Further,
$\Gamma_\sigma=4\sqrt\pi\,\alpha_\pi^2\,n_N T^{1/2} m_N^{-5/2}$
with $n_N$ the nucleon density, $m_N$ the nucleon mass, and
$\alpha_\pi\equiv(f 2m/m_\pi)^2/4\pi\approx15$ with $f\approx1$ the
pion-nucleon ``fine-structure constant.''
The function $\bar s$ is explicitly  
\begin{eqnarray}\label{eq:snd}
\bar s(x,y)&=&\frac{e^{-x/2}+e^{x/2}}{16}
\int_{|x|}^{\infty}
dt\,\,e^{-t/2}\,\,\frac{3x^2+6ty+5y^2}{3}\nonumber\\
&&{}\times\Biggl[\frac{2\sqrt{t^2-x^2}}{x^2+2ty+y^2}\nonumber\\
&&\hskip2em{}-\frac{1}{t+y}\,
\log\left(\frac{t+y+\sqrt{t^2-x^2}}{t+y-\sqrt{t^2-x^2}}\right)\Biggr],
\end{eqnarray}
where $x=\omega/T$ and pion-mass effects are included by virtue of the
parameter $y\equiv m_\pi^2/m_N T=19.4~{\rm MeV}/T$.  We show $\bar
s(x,y)$ in Fig.~\ref{fig:snd} for several values of $y$.

For temperatures between 10 and $40~{\rm MeV}$, corresponding to
$y=0.5$--2, this function is approximately constant for a large
range of $x=\omega/T$. If we approximate it by $1/2$ we may 
write $\bar S^{(1)}(\omega)\approx 2\Gamma/\omega^2$ with
$\Gamma=\Gamma_\sigma/4$ or
\begin{equation}\label{eq:perturb}
\gamma\equiv\frac{\Gamma}{T}
=\frac{\sqrt\pi\,\alpha_\pi^2\,n_N}{T^{1/2}m_N^{5/2}}
=1.24\,\rho_{14}\,T_{30}^{-1/2},
\end{equation}
where $\rho_{14}=\rho/10^{14}~{\rm g}~{\rm cm}^{-3}$ and
$T_{30}=T/30~{\rm MeV}$. Even if $\gamma$ is only a few is enough to
imply a significant cross-section suppression (Appendix~B).


\section{Cross-Section Reduction}

The reduction of the axial-current neutrino-nucleon scattering cross
section $\sigma$ by spin fluctuations was discussed in detail in
Ref.~\cite{RSS}.  With $\sigma_0$ the vacuum cross section and
$\langle\cdots\rangle$ an average over a Maxwell-Boltzmann
distribution of neutrino energies, one finds~\cite{RSS}
\begin{equation}\label{eq:suppression}
\frac{\langle\sigma\rangle}{\langle\sigma_0\rangle}
=\int_0^\infty\frac{d\omega}{2\pi}\,S(\omega)\,
\left[2+\frac{\omega}{T}+
\frac{1}{6}\,\left(\frac{\omega}{T}\right)^2\right]\,
e^{-\omega/T}.
\end{equation}
Equivalently, the amount of reduction can be expressed in the form
\begin{equation}\label{eq:suppression2}
\frac{\delta\langle\sigma\rangle}{\langle\sigma_0\rangle}
=-\int_0^\infty\frac{d\omega}{2\pi}\,S(\omega)\,G(\omega/T),
\end{equation}
where the dimensionless phase-space weight function is
\begin{equation}\label{eq:Gdef}
G(x)=1-(1+x+x^2/6)\,e^{-x}.
\end{equation}
This representation is needed to determine the suppression when the
structure function $S(\omega)$ is given by an un-resummed perturbative
expression and thus diverges at $\omega=0$.

\begin{figure}[ht]
\epsfxsize=8cm
\hbox to\hsize{\hfill\epsfbox{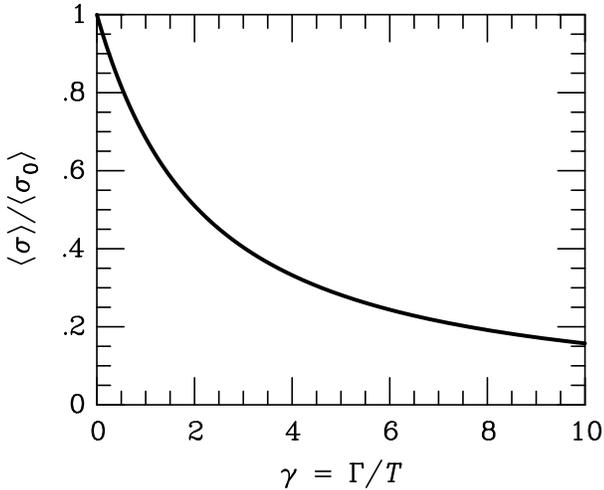}\hfill}
\smallskip
\caption{\label{fig:suppression}
Suppression of the average 
axial-current neutrino-nucleon scattering
cross section according to Eq.~(\protect\ref{eq:suppression}).}
\end{figure}

We show the cross-section suppression Eq.~(\ref{eq:suppression}) in
Fig.~\ref{fig:suppression} as a function of $\gamma=\Gamma/T$ for our
generic structure function
\begin{equation}
S(\omega)=\frac{2\Gamma}{\omega^2+\Gamma^2}\,\frac{2}{1+e^{-\omega/T}}
\end{equation}
implied by Eqs.~(\ref{eq:sbars}) and~(\ref{eq:lor}). Limiting cases
are
\begin{equation}\label{eq:limiting}
\frac{\langle\sigma\rangle}{\langle\sigma_0\rangle}
=\cases{1-A\gamma&for $\gamma\ll 1$,\cr
B/\gamma&for $\gamma\gg 1$.\cr}
\end{equation}
The coefficients are
\begin{eqnarray}\label{eq:Coefficients}
A&=&\frac{2}{\pi} \int_0^\infty dx\,
\frac{e^x-(1+x+x^2/6)}{x^2(1+e^x)}=0.438703,
\nonumber\\
B&=&\frac{2}{\pi} \int_0^\infty dx\,
\frac{2+x+x^2/6}{1+e^x}=1.59745.
\end{eqnarray}
We will need to know the value of $\gamma$ corresponding to
a given cross-section suppression 
$s=|\delta\langle\sigma\rangle/\langle\sigma_0\rangle|$. An
approximation formula, good to within about 1\%, is
\begin{equation}\label{eq:gamma_supp}
\gamma(s)=\frac{B\,(1-\sqrt{1-s})\,s}{1-s}+\frac{s}{A\sqrt{1-s}}.
\end{equation}
It is essentially an interpolation between the inversions of
the limiting functions of Eq.~(\ref{eq:limiting}). 

\begin{figure}[ht]
\epsfxsize=8cm
\hbox to\hsize{\hfill\epsfbox{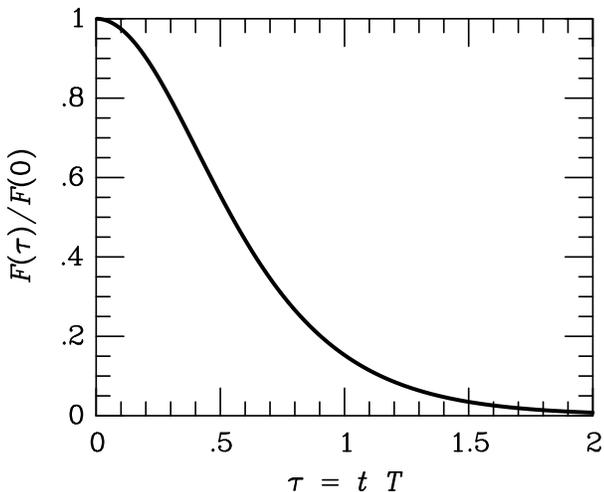}\hfill}
\smallskip
\caption{\label{fig:ftau}
Function $F(\tau)$ according to Eq.~(\protect\ref{eq:ftau}).}
\end{figure}

In our present study, the primary quantity produced by the 
numerical runs is $\bar R(t)$. Therefore, we finally express
Eq.~(\ref{eq:suppression}) in its Fourier transformed version,
\begin{equation}\label{eq:suppression3}
\frac{\langle\sigma\rangle}{\langle\sigma_0\rangle}
=\int_0^\infty d\tau\, \bar R(\tau/T)\,F(\tau)
\end{equation}
where
\begin{equation}\label{eq:ftau}
F(\tau)=\frac{2}{\pi}\int_0^\infty dx\,\cos(x\tau)\,
\frac{2+x+x^2/6}{1+e^x},
\end{equation}
shown in Fig.~\ref{fig:ftau}. It is normalized as
$\int_0^\infty d\tau\,F(\tau)=1$ and 
$F(0)=B=[\pi^2+12\log(4)+3\zeta(3)]/6\pi=1.59745$. 

\end{document}